\documentstyle[12pt]{article}
\begin{document}
\parskip 10pt plus 1pt
\title{RESOLUTION OF THE STRONG CP AND U(1)
PROBLEMS\thanks{Invited lecture at XI DAE Symposium on High
Energy Physics, Santiniketan, 1995}}
\author{
{H. Banerjee}
\\{S.N.Bose National Centre For Basic Sciences}\\
{DB-17, Sector-I, Salt Lake, Calcutta-700064, INDIA}\\
}
\baselineskip=12pt
\date{}
\maketitle

\begin{abstract}
Definition of the determinant of Euclidean Dirac operator in the
nontrivial sector of gauge fields suffers from an inherent
ambiguity. The popular Osterwalder-Schrader (OS) recipe for the
conjugate Dirac field leads to the option of a vanishing
determinant. We propose a novel representation for the conjugate
field which depends linearly on the Dirac field and yields a
nonvanishing determinant in the nontrivial sector. Physics, it
appears, chooses this second option becuase the novel
representation leads to a satisfactory resolution of two
outstanding problems, the strong CP and U(1) problems,
attributed to instanton effects.
\end{abstract}
\newpage

\section{Introduction}
The strong CP and U(1) problems are two outstanding problems in
the QCD sector of the standard model. The former consists in the
gross disagreement of theoretical estimates of electric dipole
moment of neutron (NEDM), which are invariably larger, by nearly
nine to ten orders of magnitude, than the experimental upper
limit [1]. The U(1) problem epitomises the difficulties in
formulating a theoretically consistent framework to interpret the
mass of the flavour singlet pseudoscalar meson $\eta^{'}$ [2],
which, unlike the other Goldstone bosons, is very heavy.

The genesis of both the problems is the anomaly for space-time
indepedent `global' chiral rotation of fermi (quark) fields
\begin{equation}
q(x) \rightarrow e^{i\alpha\gamma_{5}} q(x), \hspace{.5cm}
\bar{q}(x) \rightarrow \bar{q}(x) e^{i\alpha\gamma_{5}}
\end{equation}
In perturbation theory there is no trace of this `global' anomaly
[3]. A space-time indepedent chiral phase in fermion mass drops
out from all amplitudes diagram by diagram if the interactions
are chirally invariant. This is not in conflict with the ABJ
anomaly in the four divergence of the axial vector current which
arises from the triangle diagram in perturbation theory. The
`global' anomaly is just the space-time integral of the ABJ
anomaly, which, in path integral approach, corresponds to
space-time dependent `local' chiral transformations
\begin{equation}
q(x) \rightarrow e^{i\alpha(x)\gamma_{5}} q(x), \hspace{.5in}
\bar{q}(x) \rightarrow \bar{q}(x) e^{i\alpha(x)\gamma_{5}}
\end{equation}
In perturbation theory the integrand is nontrivial but the
integral vanishes.

The sine qua non of a nonvanishing integral and hence of `global'
chiral anomaly is the zero modes of Euclidean Dirac operator
which live in compactified space-time and are inaccessible in
perturbative  framework. The carrier of the virus of fermion zero
modes in popular path integral approach is identified to be the
Osterwalder-Schrader (OS) [4] recipe for the conjugate Dirac field
in Euclidean metric, viz., $\bar{q}(x)$ is indepedent of $q(x)$.
In relativistic metric the relation $\bar{q}(x) = (\gamma_{o}
q(x))^{+}$ relates the Dirac field to its conjugate. The OS
recipe, therefore, requires that the degrees of freedom are
doubled in Euclidean metric. This is neither natural nor
necessary.

We propose a prescription which is just the opposite of the OS
recipe. To be precise, we require $\bar{q}(x)$ to be antilinear
in $q(x)$ and the relationship to obey reciprocity [5]. The
representation of $\bar{q}(x)$ is unique if it is required that
$\bar{q}(x)$ has the correct chiral properties, i.e., obeys
eq.(1). The novel representation reproduces the correct two-point
correlation function, and hence, by antisymmetry, all the 2n-point
fermion correlation functions of perturbation theory [5].This
assures that the novel representation is consistent with all the
standard axioms of Euclidean field theory [6]. The point of
interest is that the fermion zero modes are evaded and the novel
representation leads to a formulation of QCD which is free from
`global' chiral anomaly. This solves the strong CP and the U(1)
problems [5].

It is remarkable that the novel representation yields, in path
integral approach, a nonvanishing determinent of the Dirac
operator $D\!\!\!\!/\,$ in the nontrivial sector $\nu \not= 0$ of gauge fields
\begin{equation}
\nu \equiv \frac{g^{2}}{16\pi^{2}} \int tr F_{\mu\nu}
\tilde{F}_{\mu\nu} d^{4}x
\end{equation}
whereas, the OS recipe gives a vanishing result. This reflects
an inherent ambiguity in the definition of the determinant of
Dirac operator in non-trivial sector. The ambiguity becomes
transparent in Weyl space where one has the option to write
Dirac determinant either as
\begin{equation}
{\em det} D\!\!\!\!/\, = {\em det} (-DD^{+}) {\em
or}\hspace{.1in} {\em det} (-D^{+}D)
\end{equation}
in terms of Weyl operators $D, D^{+}$ defined through
\begin{eqnarray}
D\!\!\!\!/\, &\equiv &\gamma_{\mu} (i\partial_{\mu} - g A_{\mu}) \nonumber
\\
   &=      &(\begin{array}{cc} 0        &D\\
			       D^{+}    &0\end{array})
\end{eqnarray}
In nontrivial sector $\nu \not= 0$ one of the options in (4)
vanishes while the other does not. This follows from the index theorem
\begin{eqnarray}
\nu &= &{\em dim} {\em ker} (DD^{+}) - {\em dim} {\em ker} (D^{+}D)
\nonumber \\
    &= &n_{+} - n_{-}
\end{eqnarray}
and the theorem that there are no `wrong chirality' solutions of
Dirac operator [7], i.e., for $\nu \geq 0$ $(\nu \leq 0)$, the
number of normalisable negative (positive) chirality solution
$n_{-} (n_{+})$ is zero. The novel representation chooses the
nonvanishing option for the Dirac determinant while the OS
recipe leads to the vanishing option in nontrivial sector.

Resolution of the strong CP and the U(1) problems, and hence, it
appears, physics chooses the option of the novel representation.

\section{Global Chiral Anomaly}
In Euclidean metric the QCD action is given by
\begin{equation}
S_{QCD} = S_{G} + S_{F} + \theta_{ew} \Delta^{'}S +
\gamma_{QCD} \Delta S
\end{equation}
where $S_{F}$ is the fermionic piece
\begin{equation}
S_{F} \equiv \int \bar{q}(x) (D\!\!\!\!/\, - iM) q(x) d^{4}x
\end{equation}
and $S_{G}$ is the gluon action. The (light) quarks have three
flavours. For convenience, we assume the mass matrix M to be
diagonal in flavour space and suppress the flavour indices.

The two terms $\Delta S$ and $\Delta^{'}S$ are potential sources
of CP violation. While the $\theta_{QCD}$ term is attributed to
the topological structure of the QCD vacuum
\begin{equation}
\Delta S \equiv \frac{g^{2}}{16\pi^{2}} \int tr F_{\mu\nu}
\bar{F}_{\mu\nu} d^{4}x
\end{equation}
the chiral phase $\theta_{ew}$ in quark mass
\begin{equation}
\Delta^{'}S \equiv \int \bar{q}(x) M\gamma_{5} q(x) d^{4}x
\end{equation}
arises from Higgs interactions in the electroweak sector. In
compactified space-time $\Delta S$ assumes integral values $\nu$.

The degrees of freedom of $q(x)$ are the Grassmann gennators
which appear as coefficients in the expansion of $q(x)$ in a
complete set of basis functions.For convenience, the orthonormal
set of eigenfunctions of the Dirac operator are chosen as basis
functions,
\begin{equation}
q(x) = \sum_{r} (a_{r} + a_{-r} \gamma_{5}) \varphi_{r}(x) +
\sum_{i} a_{oi} \varphi_{oi}(x)
\end{equation}
where the normalised eigenfunctions obey the equations,
\begin{eqnarray}
D\!\!\!\!/\, \varphi_{r}(x) = \lambda_{r} \varphi_{r} (x) &, &\hspace{.5in}
D\!\!\!\!/\, \gamma_{5}\varphi_{r}(x) = -
\lambda_{r}\gamma_{5}\varphi_{r}(x) \nonumber \\
D\!\!\!\!/\, \varphi_{oi}(x) = 0 &, &\hspace{.5in} \gamma_{5}\varphi_{oi}(x) =
\epsilon_{i}\varphi_{oi}
\end{eqnarray}
The zero eigenmodes $\varphi_{oi}$ have definite chiralities,
positive $(\epsilon_{i} = 1)$ or negative $(\epsilon_{i} = -1)$.
The OS recipe is implemented by choosing an independent set of
Grassmann generators for $\bar{q}(x)$
\begin{equation}
\bar{q}(x) = \sum (b_{r} + b_{-r} \gamma_{5}) \varphi_{r}(x) +
\sum b_{oi}\varphi_{oi}(x)
\end{equation}

The Jacobian of the measure in the fermionic partition function $Z_{F}$
\begin{equation}
Z_{F} \equiv \int d\mu {\em exp} [- S_{F}]
\end{equation}
has the form
\begin{equation}
ln J(\alpha) = - 2i\alpha \int A(x)d^{4}x
\end{equation}
for `global' chiral rotation (1). The integrand A(x) is
identified as the `local' chiral anomaly.

The measure corresponding to OS recipe for $\bar{q}(x)$ is given
by
$$d\mu^{I} = \Pi_{r} da_{r} db_{r} \Pi da_{oi} db_{oi}$$
The `local' chiral anomaly for this measure was obtained by
Fujikawa [8],
\begin{eqnarray}
A^{I}(x) &=
%% FOLLOWING LINE CANNOT BE BROKEN BEFORE 80 CHAR
&2\sum\varphi_{r}^{+}(x)\gamma_{5}\varphi_{r}(x)+\sum\epsilon_{i}\varphi^{+}_{oi}(x)\varphi_{oi}(x)\nonumber\\
	 &= &\frac{g^{2}}{16\pi^{2}} trF_{\mu\nu}(x) \bar{F}_{\mu\nu}(x)
\end{eqnarray}

Nonzero eigenmodes drop out because of orthogonality of
$\varphi_{r}(x)$ and $\gamma_{5}\varphi_{r}(x)$ and only zero
modes of $A^{I}(x)$ survive in the integral (15) for the Jacobian
\begin{eqnarray}
ln J^{I}(\alpha) &= &-2i\alpha(n_{+} - n_{-}) \nonumber \\
		 &= &-2i\alpha\frac{g^{2}}{16\pi^{2}} \int tr
F_{\mu\nu}\bar{F}_{\mu\nu}d^{4}x
\end{eqnarray}
This means that under `global' chiral rotation QCD action
changes according to the formula
\begin{equation}
S_{QCD} \rightarrow S^{I}_{QCD}(\alpha) = S_{G} + S_{F} +
(\theta_{ew} + 2\alpha) \Delta^{'}S + (\theta_{QCD} -
6\alpha)\Delta S
\end{equation}

In effective Lagrangians for chiral models there is no scope for
a nontrivial Jacobian. The global chiral U(1) anomaly (17) in
underlying QCD can, therefore, be reproduced in effective
Lagrangians through an `anomaly term' which breaks chiral
symmetry explicitly. A popular representation of the anomaly
term is [9]
\begin{equation}
\Delta S^{I}_{eff} = - m^{2}_{\eta^{'}} f^{2}_{\pi} \int
[tr ln(\frac{U}{U^{+}}) - \theta_{QCD}]^{2}d^{4}x
\end{equation}
where $f_{\pi}$ and $m^{'}_{\eta}$ are respectively the pion
decay constant and the mass of the flavour singlet Goldstone
boson. The meson matrix U transforms as $U \rightarrow
Ue^{i\alpha}$ under global chiral rotation (1). The problems
with this `anomaly term' are (a) its chiral variation depends
explicitly on $\theta_{QCD}$, and (b) its second order variation
does not vanish. Neither of these properties hold in the
underlying QCD. This is the crux of the controversy between 't
Hooft and Crewther [2], and the reason why the popular
resolution (19) of the U(1) problem is regarded as unsatisfactory.

The transformation law (18) shows that neither $\theta_{ew}$ nor
$\theta_{QCD}$ can be physical. Only the chirally invariant
combination $\bar{\theta} \equiv (\theta_{QCD} + 3\theta_{ew})$
can appear in CP violating effects. Theoretical estimates in
various chiral models suggests [1] NEDM in the range
$$d_{n} \approx \bar{\theta} \times 10^{-15 \pm 1} e.cm$$
Experimental upper limit $\vert d_{n}\vert < 10^{-25} e.cm$,
therefore, puts a stringent constraint $\bar{\theta} <
10^{-10}$. This is the crux of the strong CP problems. The
strong CP problem is, therefore, a serious problem of fine
tuning. Two parameters $\theta_{QCD}$ and $3\theta_{ew}$ which
are of completely different origins in the standard model must
be so fine tuned as to cancel each other completely.

\section{Novel representation of Euclidean Dirac fermion}
We start from the ansatz which is just the opposite of the OS
recipe, i.e., we assume that in Euclidean metric the conjugate
field $\chi(x) \equiv [\bar{q}(x)]^{+}$ is linear in $q(x)$.
This means that the Grassmann generators defining the degrees of
freedom of $\chi(x)$ are a subset of the generators $\{a_{r},
a_{oi}\}$ appearingin $q(x)$. The resulting representation of
$\chi(x)$ is unique, modulo an overall sign, if one requires
that, (a) chiral charge of $\chi(x)$ is opposite to that of
$q(x)$, i.e., if $q(x) \rightarrow e^{i\alpha\gamma_{5}} q(x)$
then $\chi(x) \rightarrow e^{-i\alpha\gamma_{5}} q(x)$, and
(b) the linear relation obeys reciprocity,
\begin{equation}
\chi(x) = \sum_{r}[a_{r} - a_{-r} \gamma_{5}]\varphi_{r}(x)
\end{equation}
The crucial point to note is that $\chi(x)$ cannot contain the
zero mode generators $a_{oi}$ which transform as $a_{oi}
\rightarrow a_{oi} e^{i\epsilon_{i}\alpha}$ under chiral
rotation (1). This will be in conflict with the chiral charge of
$\chi(x)$. As a result, the fermion action $S_{F}$ is devoid of
the zero mode generators
\begin{eqnarray}
S_{F} &= &\int \chi^{+}(x) (D\!\!\!\!/\, - iM) q(x)d^{4}x \nonumber \\
      &= &\sum_{r} [\lambda_{r}(a^{*}_{r}a_{r} +
a^{*}_{-r}a_{-r}) - iM (a^{*}_{r}a_{r} - a^{*}_{-r}a_{-r})]
\end{eqnarray}
The measure appropriate for this action
$$d\mu = \Pi_{r} da^{*}_{r} da_{r} da^{*}_{-r} da_{-r}$$
leads to the partition function
\begin{equation}
Z_{F} = \Pi_{\lambda_{r}>o} (\lambda_{r}^{2} + M^{2})
\end{equation}
whose chiral limit (M=0) does not vanish in the nontrivial
sector $(\nu \not= 0)$ of gauge fields.

The two-point correlation function, obtained in the usual path
integral approach, coincides with the familiar formula
\begin{equation}
< q(x) \bar{q}(y) > = < x\vert\frac{1}{D\!\!\!\!/\, - iM}\vert y > -
\frac{\sum_{i}\varphi_{oi}(x)\varphi^{+}_{oi}(y)}{-iM}
\end{equation}
except that the zero mode contributions are subtracted out. In
the limit of zero coupling $g = 0$, the momentum representation
of the correlation function coincides with the Wick-rotated
relativistic Feynman propagator
\begin{equation}
< q(x) \bar{q}(y) >_{g=0} = \frac{1}{(2\pi)^{4}} \int d^{4}p
\frac{p\!\!\!/ + iM}{p^{2} + M^{2}} e^{-ip(x-y)}
\end{equation}
The remaining 2n-point correlation functions follow, in path
integral framework, from the anticommutation of $q(x)$ and $\bar{q}(y)$
\begin{equation}
< q(x_{1}) ... q(x_{m}) \bar{q}(y_{1}) ... \bar{q}(y_{n}) > =
\delta_{mn} {\em det} [< q(x_{i}) \bar{q}(y_{i})]
\end{equation}

The `local' chiral anomaly is given by an expression analogous
to that in the OS formulation (16) except that zero modes are
excluded from the sum on the right hand side,
\begin{eqnarray}
A^{II(x)} &= &2\sum_{\lambda_{r}>0} \varphi^{+}_{r}(x) \gamma_{5}
\varphi_{r}(x) \nonumber \\
	  &= &\frac{g^{2}}{16\pi^{2}} \int tr F_{\mu\nu}
\bar{F}_{\mu\nu} d^{4}x - \epsilon_{i}\varphi^{+}_{oi}(x) \varphi_{oi}(x)
\end{eqnarray}
Thus the four divergence of the U(1) axial current has a
nontrivial anomaly $A^{II}(x)$
\begin{equation}
\partial_{\mu} J_{\mu 5} (x) = 2\bar{q}(x) M \gamma_{5} q(x) + A^{II}(x)
\end{equation}
which coincides, as it must, with the perturbative (absence of
zero modes) ABJ anomaly. However, the `global' chiral anomaly vanishes
\begin{eqnarray}
ln J^{II}(\alpha) &= &- 2i\alpha \int A^{II}(x) d^{4}x \nonumber
\\
&= &0
\end{eqnarray}
and our desired goal is achieved. The vanishing is the direct
consequence of the orthogonality of nonzero eigenmodes
$\varphi_{r}(x)$ and $\gamma_{5}\varphi_{r}(x)$, or, if one
prefers, the index theorem.

The novel representation (20) is equivalent to the functional relation
\begin{equation}
\chi(x) = \frac{1}{[D\!\!\!\!/\, ^{2}]^{\frac{1}{2}}}D\!\!\!\!/\, q(x)
\end{equation}
which is nonlocal. This is not a disability in Euclidean field
theory. Locality is a field theoretic axiom only in rleativistic
metric. This translates in Euclidean field theory into the axiom
of (anti) symmetry of correlation functions under permutations
[6]. The fact that all the correlation functions are reproduced
correctly through the eqs.(24) and (25) assures us that the
novel representation (20, 29) is consistent not only with the
axiom of symmetry but with the other axioms of Euclidean field
theory, e.g., reflection positivity, cluster decomposition etc.,
as well.

\section{Resolution of strong CP and U(1) problems}
In the scenario corresponding to the novel representation (20,
29) the QCD action is invariant under `global' chiral rotation
in the chiral (M = 0) limit
\begin{equation}
S_{QCD} \rightarrow S^{II}_{QCD}(\alpha) = S_{G} + S_{F} +
(\theta_{ew} + 2\alpha) Delta^{'}S + \theta_{QCD} \Delta S
\end{equation}
In effective Lagrangians, the `anomaly' term, which is invariant
under `global' chiral rotation but reproduces the ABJ anomaly in
axial Ward identity, is easily constructed
\begin{equation}
\Delta S^{II}_{eff} = - m^{2}_{\eta^{'}} f^{2}_{\pi} \int [tr ln
(\frac{U}{U^{+}}) - < tr ln (\frac{U}{U^{+}})>]^{2} d^{4}x
\end{equation}
The controversial features of the popular construction (19) thus
disappear and the U(1) problem satisfactorily resolved [5].

The law of transformation (30) shows that the parameter
$\theta_{QCD}$, the coefficient of $\Delta S$, is invariant,
while $\theta_{ew}$ is unphysical and can be eliminated
trivially by `global' chiral rotation (1). There is no longer
any problem of fine tuning and CP symmetry of QCD is simply
ensured through the `natural' choice $\theta_{QCD} = 0$ [5].

We conclude that the strong CP and the U(1) problems both are
legacies of the OS recipe. The alternative scenario of QCD with
the novel representation (20, 29) for the conjugate Dirac field
is not afflicted with these blemishes [5].

\end{document}